\documentclass[twocolumn,showpacs,preprintnumbers,amsmath,amssymb]{revtex4}

\usepackage{graphicx}
\usepackage{rotating}
\usepackage{dcolumn}
\usepackage{bm}

\begin{document}


\title{Baroclinic equivalent and nonequivalent barotropic modes for rotating stratified flows}
\author{M. Jia$^{1}$ and S. Y. Lou$^{1,2,3}$}
\affiliation{ \it \small $^{1}$Department of Physics, Shanghai Jiao
Tong University, Shanghai, 200240, China\\
 \small $^{2}$Faculty of Science, Ningbo University, Ningbo,
315211, China\\
 \small $^3$School of Mathematics, Fudan University, Shanghai,
200433, China }

\date{\today}

\begin{abstract}
In strictly speaking, all the natural phenomena on the earth should
be treated under rotating coordinate. The existence of baroclinic
nonequivalent barotropic laminar solution for rotating fluids is
still open though the laminar solutions for the irrotational fluid
had been well studied. In this letter, all the possible equivalent
barotropic (EB) laminar solution are firstly explored and all the
possible baroclinic non-EB elliptic circulations and hyperbolic
laminar modes are discovered. The baroclinic EB circulations
(including the vortex streets and hurricane like vortices) possess
rich structures because either the arbitrary solutions of arbitrary
nonlinear Poison equations can be used or an arbitrary
two-dimensional stream function is revealed. The discovery of the
baroclinic non-EB modes disproves a known conjecture. The results
may be broadly applied in atmospheric and oceanic dynamics, plasma
physics, astrophysics and so on.
\end{abstract}
 \maketitle

\em 1. Introduction. \rm It is known that both the planetary
rotations and stable vertical density stratification are important
for the fluid motions in atmospheres and oceans. The effect of
rotation and stratification are the most important features that
distinguish fluid flow in the atmosphere and ocean. The flows in the
rotating stratified fluids exhibit rich phenomena especially on the
circulation vortices \cite{Science,Nature} like the Jupiter's Red
Spot, tropospheric cyclones, hurricanes\cite{hurricane,Lou},
tornados, stratospheric polar vortices, oceanic Gulf Stream rings,
atmospheric blockings which are mainly responsible for many kinds of
meteorological disasters such as the floods, droughts and snowstorms
etc. \cite{Luo,Huang}

For the usual irrotational fluid, the laminar solution is studied
quite well. However, for the rotational fluid, there are many open
problems on laminar solutions.

Recently, a steady baroclinic laminar model
\begin{eqnarray}
&&u u_x+v u_y-f v=-p_x,\ u v_x+v v_y+f u=-p_y,\nonumber\\
&&p_z=-\rho, \quad u_x+v_y=0, 
\quad u \rho_x+v \rho_y=0, \label{eq5}
\end{eqnarray}
where $f$ is the Coriolis parameter, $p$ is the pressure
perturbations divided by a mean density $\rho_0$ and $\rho$ is the
density perturbation scaled by $\rho_0/g$, $u$ and $v$ are
horizontal velocities while the vertical velocity $w$ has been
dropped out because its weakness, is developed as the late-time
equilibrium state in the free decay of rotating stratified.

To derive the model (\ref{eq5}), the author has hypothesized that
the formation mechanism for coherent structures in rotating
stratified flows is fundamentally baroclinic. However, to find exact
baroclinic solutions in fluid dynamics is very difficult and there
is little progress in this direction. In Ref. \cite{sun}, one type
of special barotropic tilting vortex solution and four special types
of baroclinic equivalent-barotropic (EB) are obtained. Basing on the
fact that all the known solutions are either
barotropic or baroclinic EB, a conjecture is proposed.\\
 {\bf Conjecture:} \em{Baroclinic solutions to \eqref{eq5}
are always EB.}\rm

Now, important questions are: How to find possible baroclinic modes
of \eqref{eq5}? Is the conjecture correct?\\
\em 2. Baroclinic EB modes. \rm Here, we try to find \em all \rm the
possible Baroclinic EB modes of the baroclinic laminar model
(\ref{eq5}).

From the incompressible condition, $u_x+v_y=0$, we can introduce
stream function $\psi$ as
\begin{eqnarray}
u=-\psi_y,\ v=\psi_x.\label{uv}
\end{eqnarray}
After introducing the stream function as in \eqref{uv}, five
equations shown in \eqref{eq5} are reduced to two equations for the
single function $\psi$ ($K\equiv \frac12\psi_x^2+\frac12\psi_y^2,\
\zeta\equiv\psi_{xx}+\psi_{yy}$, $J(a,b)\equiv a_xb_y-a_yb_x$)
\begin{eqnarray}
&&J(\psi,K_z)-(\zeta+f)J(\psi,\psi_{z})=0,\quad
\label{psi1}\\
&& J(\psi,\zeta)=0.\label{psi2}
\end{eqnarray}
Eq. \eqref{psi1} is just the last equation of \eqref{eq5} while  Eq.
\eqref{psi2} is the consistent condition of the first two equations
of \eqref{eq5}, i.e., $p_{xy}-p_{yx}=0$. Whence the stream function
$\psi$ is solved out from \eqref{psi1} and \eqref{psi2}, the
velocity components is obtained immediately from \eqref{uv}, the
pressure can be solved out from the consistent equations, the first
two equations of \eqref{eq5} while the density is only a simple
differentiation of the pressure with respect to $z$.

\em Definitions: \rm  The fluid is barotropic if density is a
function of pressure only, that is, isobaric surfaces and isopycnal
surfaces coincide; otherwise, the fluid is baroclinic. A baroclinic
flow is EB if the streamlines on each plane align vertically or,
equivalently, if the horizontal velocity vector does not change
direction vertically. More clearly, the fluid is barotropic iff the
pressure of \eqref{eq5} is a function of $z+h$ with an arbitrary
function $h\equiv h(x,y)$ while the fluid is called baroclinic EB of
\eqref{eq5} iff
\begin{eqnarray}
\psi_x=F \psi_y\label{EB}
\end{eqnarray}
for arbitrary $F\equiv F(x,y)$.

From \eqref{EB} and \eqref{psi1}, it is easy to find that the only
two possible cases of baroclinic EB, (A) $\psi_{yz}=0$ and (B) $F_y+FF_x=0$.\\
\em (A) Baroclinic EB with an arbitrary nonlinear Poison flow. \rm
For the $\psi_{yz}=0$ case, we have the stream function
\begin{eqnarray}
\psi=\phi(x,y)+\psi_0(z),\label{strem}
\end{eqnarray}
with $\psi_0(z)$ being an arbitrary function of $z$ while
$\phi\equiv \phi(x,y)$ is a solution of an arbitrary nonlinear
Poison equation
\begin{eqnarray}
\phi_{xx}+\phi_{yy}=g(\phi),\label{ps}
\end{eqnarray}
where $g(\phi)$ is an arbitrary function of $\phi$. Whence the
Poison equation \eqref{ps} is solved, the other quantities can
easily be found. The results read
\begin{eqnarray}
&&u=-\phi_y,\ v=\phi_x,\label{uv1}\\
&&p=\frac12\phi_y^2+\int\phi_x\phi_{yy}\mbox{\rm
dx}+f\phi+\phi_0(y)+p_0(z),
\end{eqnarray}
where $p_0(z)$ is an arbitrary function of $z$ while $\phi_0(y)$
should be appropriately fixed such that the first two equations of
\eqref{eq5} are compatible.
\input epsf
\begin{figure}
\epsfxsize=7cm\epsfysize=5cm\epsfbox{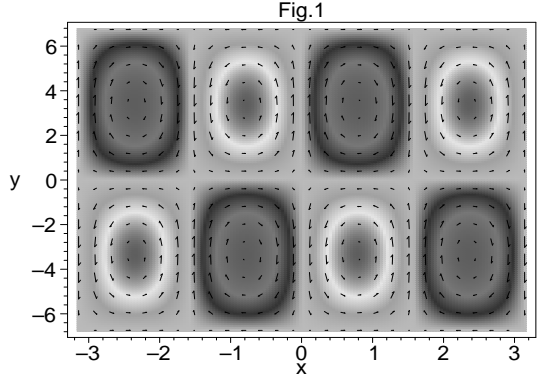} \caption{The
density plot of the vortex street solution \eqref{EB1} and the
vector field plot of the corresponding velocity field expressed by
\eqref{ru1} and \eqref{rv1} with the parameter selections
$a=\frac18,\ b=2,\ c=\frac12$.}
\end{figure}

In Fig. 1, a special vortex street solution $((m,n)\equiv
\left(a\frac{c}{b},\ a\frac{b}{c}\right))$,
\begin{eqnarray}
&&\psi=\phi=4\arctan\left(a\mbox{\rm sn}(bx,m)\mbox{\rm
sn}(cy,n)\right),\ \label{EB1}
\end{eqnarray}
where $a,\ b$ and $c$ are constants while $\mbox{\rm sn}(bx,m)$ is
the standard Jacobi elliptic function with modula $m$,  is shown
with the parameter selections $a=\frac18,\ b=2,\ c=\frac12$.

Corresponding to the solution \eqref{EB1}, the arbitrary function of
Poison equation is fixed as
\begin{eqnarray}
&&g(\phi)=-(b^2+c^2)(1+a^2)\sin(\phi),\label{gp1}
\end{eqnarray}
and the other physical quantities are
\begin{eqnarray}
&&u=-\frac{4ac\mbox{sn}(bx,m)\mbox{cn}(cy,n)\mbox{dn}(cy,n)}
{1+a^2\mbox{sn}^2(bx,m)\mbox{sn}^2(cy,n)},\label{ru1}\\
&&v=\frac{4ab\mbox{cn}(bx,m)\mbox{dn}(bx,m)\mbox{sn}(cy,n)}
{1+a^2\mbox{sn}^2(bx,m)\mbox{sn}^2(cy,n)},\label{rv1}\\
&&p=f\phi-8b^2\frac{c^2\mbox{sn}^2(bx,m)+b^2\mbox{sn}^2(cy,n)}
{1+a^2\mbox{sn}^2(bx,m)\mbox{sn}^2(cy,n)}+g,\label{rp1}
\end{eqnarray}
where $g\equiv g(z)$ is an arbitrary function of $z$.\\
 \em (B) Baroclinic EB symmetric circulations. \rm For the
$F_y+FF_x=0$ case, it is straightforward to prove that the only
possible modes are
\begin{eqnarray}
&&\psi=\psi_0,\quad r\equiv c_1(x^2+y^2)+c_2x+c_3y,\label{psicase2}\\
&&u=-\psi_{0r}(2c_1y+c_3),\ v=\psi_{0r}(2c_1x+c_2),\label{uv2}\\
&&p=2c_1\int \psi_{0r}^2\mbox{\rm dr} +f\psi_0+p_0(z),
\end{eqnarray}
where $c_1,\ c_2$ and $c_3$ are arbitrary constants while $p_0\equiv
p_0(z)$ and $\psi_0\equiv\psi_0(r,\ z)$ are arbitrary functions of
the indicated variables.

It is clear that there exist abundant symmetric circulation modes
($c_1\neq 0$ in \eqref{psicase2}) and jet modes ($c_1=0$ in
\eqref{psicase2}) because the stream function is an arbitrary
function of two variables $r$ and $z$. The richness of the symmetric
circulations for the rotational fluids is natural as one had
observed in both the oceans and the atmosphere. Actually, both in
the atmosphere and in the oceans, there are also many kinds of
nonsymmetric circulations.

\input epsf
\begin{figure}
\epsfxsize=6cm\epsfysize=5cm\epsfbox{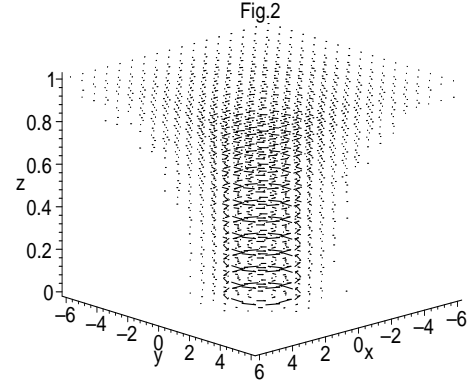} \caption{A
3-dimensional vortex solution described by the vector velocity field
\eqref{ru2}-\eqref{rv2}.}
\end{figure}

In Fig. 2, a special second type of baroclinic symmetric EB mode is
plotted for the velocity field ($r\equiv x^2+y^2-2$)
\begin{eqnarray}
&&u=2(z-1)y\mbox{\rm sech}[(1-z)r]\mbox{\rm sech}(1-z),\label{ru2}\\
&& v=2(1-z)x\mbox{\rm sech}[(1-z)r]\mbox{\rm sech}(1-z)\label{rv2}
\end{eqnarray}
which is related to the stream function solution \eqref{psicase2}
\begin{eqnarray}
&&\psi=\mbox{\rm sech}(1-z)\arctan\left\{\mbox{\rm
sinh}[(1-z)r]\right\}.\label{rpsi2}
\end{eqnarray}
Correspondingly, the pressure has the form
\begin{eqnarray}
p&=&2(1-z)\tanh[(1-z)r]\mbox{\rm sech}^2[(z-1)r]
\nonumber\\
&&-2f\mbox{\rm
sech}(1-z)\arctan\left\{\exp[(z-1)r]\right\}+p_0,\label{rp2}
\end{eqnarray}
where $p_0$ is still an arbitrary function of $z$.
\input epsf
\begin{figure}
\epsfxsize=6cm\epsfysize=5cm\epsfbox{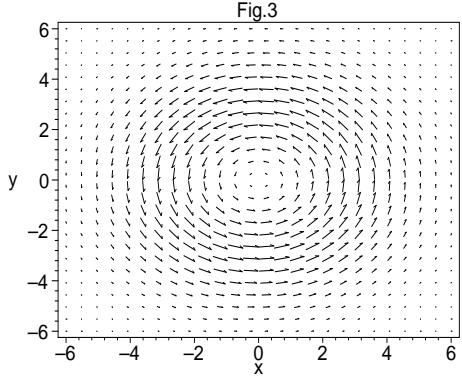} \caption{The
hurricane like structure which is the bird's eye view of Fig. 2.}
\end{figure}
The overlooking form of Fig. 2 has the form of Fig. 3 which
exhibits a hurricane like circulation form with a hurricane eye. \\
 \em 3. Baroclinic non-EB elliptic and hyperbolic modes. \em To
find a nonsymmetric circulation, we restrict ourselves to find
elliptic or hyperbolic modes of \eqref{eq5}. For an elliptic or
hyperbolic mode is defined as its stream lines are elliptic and/or
hyperbolic curves. In other words, the stream function $\psi$ has
the form
\begin{eqnarray}
\psi=\psi(a_1(z)(x-x_0(z))^2+a_2(z)(y-y_0(z))^2,z),\label{stream}
\end{eqnarray}
with $a_1(z)a_2(z)>0$ for elliptic and $a_1(z)a_2(z)<0$ for
hyperbolic modes.

Substitute \eqref{stream} into \eqref{psi2}, one can easily find
\begin{eqnarray}
\psi_\xi\psi_{\xi\xi}[a_2(z)-a_1(z)][y-y_0(z)][x-x_0(z)]=0,
\label{10}
\end{eqnarray}
where $\xi\equiv a_1(z)(x-x_0(z))^2+a_2(z)(y-y_0(z))^2$.

 From \eqref{10}, we know that the only case is
$\psi_{\xi\xi}=0$ for nonsymmetric ($a_2(z)\neq a_1(z)$) modes,
i.e.,
\begin{eqnarray}
\psi=a_1(z)(x-x_0(z))^2+a_2(z)(y-y_0(z))^2+\psi_0(z)\label{stream1}
\end{eqnarray}
which lead to that the equation \eqref{psi1} is correct only for the
two nontrivial cases:\\
\em Case 1. Baroclinic elliptic or hyperbolic non-EB modes with
rotational shape as the height $z$ changes:\rm
\begin{eqnarray}
&&\psi_{\pm}=\pm \frac1{2h}g_{\pm}\eta_{\pm}^2+\psi_0\label{stream11}\\
&&u_{\pm}=\frac1{h}g_{\pm}(y-y_0),\ v=g_{\pm}{h}(x-x_0),\label{uv1}\\
&& p_{\pm}=p_0+\frac12g_{\pm}\left(g_{\pm}\eta^2\pm
f\frac{\eta_{\pm}^2}{h}\right),\label{rp1}
\end{eqnarray}
where $h,\ \psi_0$ and $p_0$ are arbitrary functions of $z$,
$\{x_0,\ y_0,\ c_1\}$ are arbitrary constants while
$\eta_{\pm}^2\equiv (y-y_0)^2\pm h^2(x-x_0)^2, \ \eta^2\equiv
(x-x_0)^2+(y-y_0)^2,\ g_+\equiv c_1-\mbox{\rm arctanh} (h),\
g_-\equiv c_1-\mbox{\rm arctan} (h)$. The upper sign is related to
the baroclinic elliptic circulation while the lower sign corresponds
to the baroclinic hyperbolic wave case.

Fig. 4 displays a special structure for a baroclinic elliptic
circulation with the velocity field
\begin{eqnarray}
&&u=\frac{y}{10}(1+z^2)\left(\mbox{\rm arctanh} \frac1{1+z^2}-\frac32\right),\label{ru3}\\
&&v=\frac{x}{10(1+z^2)}\left(\frac32-\mbox{\rm arctanh}
\frac1{1+z^2}\right)\label{rv3}
\end{eqnarray}
which corresponds to the selections $h=\frac1{1+z^2},\ c_1=\frac32,\
x_0=y_0=0,\ f=\frac1{10}$ in \eqref{uv1}.

From \eqref{stream11}-\eqref{rp1}, we find that the baroclinic
elliptic circulation possesses some interesting properties. (i) The
circulation center is independent of the height $z$. (ii) The length
of the elliptic axes are changeable as $z$ and then the circulation
shape is rotated as $z$ changes. (iii) All the quantities, the
stream function, the velocity field and the pressure and density,
possess elliptic distributions. (iv) The rotation direction of the
vortex may be changeable if $g_+=c_1-\mbox{\rm arctanh} (h)=0$ has a
solution. Otherwise the rotation direction of the vortex will be
independent of the height variable $z$.
\input epsf
\begin{figure}
\epsfxsize=6cm\epsfysize=5cm\epsfbox{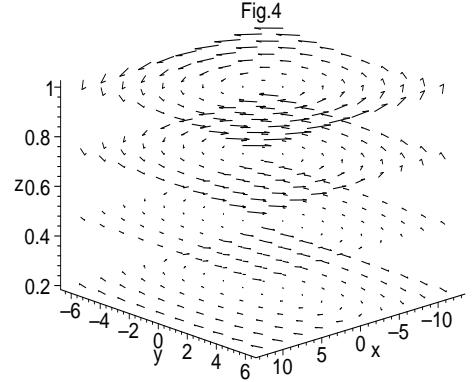}
\caption{Baroclinic elliptic circulation for the vector velocity
field described by \eqref{ru3}-\eqref{rv3}.}
\end{figure}
\\
\em Case 2. Baroclinic elliptic or hyperbolic non-EB mode with skew
center.\rm
\begin{eqnarray}
&&\psi=c(x-x_0(z))^2-\frac f2(y-y_0)^2+\psi_0(z),\label{stream13}\\
&&u=f(y-y_0),\ v=2c(x-x_0(z)),\label{uv3}\\
&& p=p_0(z)+\frac12f(2c+f)(y-y_0)^2,\label{rp3}
\end{eqnarray}
where $x_0(z),\ \psi_0(z)$ and $p_0(z)$ are arbitrary functions of
$z$ and $\ y_0$ and $c$ are arbitrary constants. If we make the
exchanges of $\{x,\ x_0,\ u\}\leftrightarrow \{y,\ y_0,\ -v\}$ in
\eqref{stream13}-\eqref{rp3}, the solution is still correct. When
$c<0$, the solution \eqref{stream13}--\eqref{rp3} is related to the
baroclinic elliptic non-EB circulation while the baroclinic
hyperbolic non-EB mode governed by $c>0$.

Fig. 5 shows us a special type of structures of
\eqref{stream13}-\eqref{rp3} for the vector velocity field with the
parameter and function selections
\begin{eqnarray}
f=\frac12,\ c=-1,\ y_0=0,\ x_0(z)=z^2. \label{para5}
\end{eqnarray}
\input epsf
\begin{figure}
\epsfxsize=6cm\epsfysize=5cm\epsfbox{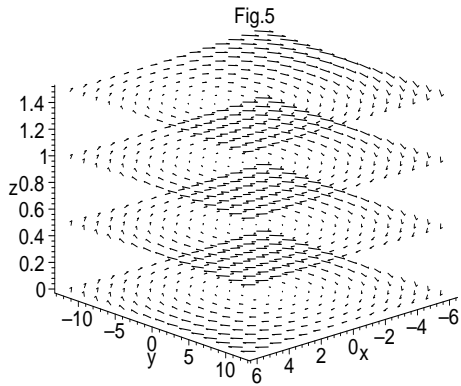} \caption{The
structure of the baroclinic elliptic circulation with skew center
for the velocity field \eqref{uv3} and the parameter selection
\eqref{para5}.}
\end{figure}
Different from the first type of baroclinic non-EB modes shown by
\eqref{stream11}--\eqref{rp1}, $\{x_0,\ y_0\}$, the center of the
second type of the baroclinic non-EB circulation
\eqref{stream13}--\eqref{rp3}, is changeable as the height changes
while the length of axes of the circulation is independent of $z$.
That means the circulation has fixed shape with skew center. On the
other hand, the pressure and the density distributions have no
circulation structure though the stream function and the velocity
field do. The rotation direction of the
vortex is always independent of the height variable.\\
\em 4. Summary and discussions. \rm In summary, the fluid systems on
the earth such as the oceans and atmosphere have to be studied under
the rotational coordinates, the laminar modes for the rotating
stratified flows have not yet been studied well though the related
topics are successfully studied for the usual irrotational fluid
systems.

In this letter, we have studied some types of steady baroclinic
equivalent and nonequivalent barotropic modes for rotating
stratified flows for the Lilly-Sun model \cite{sun,Lilly}. Usually,
to find some exact baroclinic solutions for the rotating fluids is
very difficult and only some quite special solutions are found.
Using the proper definitions, all the possible baroclinic EB models
are obtained. The first type of baroclinic EB modes are determined
up to an arbitrary symmetric Poison equation which allowed us to get
infinitely many exact solutions including vortex street like
solutions. The second type of baroclinic EB solutions are quite free
because of the existence of an arbitrary stream function with two
arbitrary variables. This kind of solutions exhibit rich structures
of the jet modes and the symmetric circulations including some
hurricane like structures.

In addition to the abundant symmetric circulations all the possible
(two types of) elliptic circulations and/or hyperbolic modes are
found. It is interesting that the discovery of this kind of
solutions disproves Sun's conjecture \cite{sun} because they are
baroclinic and non-EB modes. For the first type of nonsymmetric
circulations, the length of the elliptic axes, the rotation
directions, may be changed with respect to the height $z$ while
their circulation center is independent of $z$. For the second type
of nonsymmetric circulations, the length of the elliptic axes and
the rotation directions are independent of $z$ while the circulation
center may skewed as the change of $z$.

The studies on the vortex solutions of the fluid systems are useful
not only in fluid (including atmospheric and oceanic) dynamics but
also in many other physical fields including the condense matter,
plasma physics, nuclear physics, astrophysics and cosmology
\cite{NSP,NSP1,NSP2}.

In this paper, rich types of steady modes for rotating stratified
flows have been obtained because of the entrance of some arbitrary
functions. Though there are also abundant circulations in the nature
such as the hurricanes, tornados, ocean circulations etc., we hope
that experimental scientists will find some of exact modes mentioned
in this letter.

\section*{Acknowledgement}
The authors are indebt to thanks Dr. X. Y. Tang, Prof. C. Sun and
Prof. F. Huang for their helpful discussions. The work was supported
by NNSFC (Nos. 10475055, 10601033 and 40305009) and NBRPC (973
Program 2007CB814800).

\end{document}